\documentclass[12pt]{article}

\usepackage{epsfig}

\begin{document}

\thispagestyle{empty}
\noindent\hspace*{\fill}  FAU-TP3-01/12 \\
\noindent\hspace*{\fill}  hep-th/0112105 \\
\noindent\hspace*{\fill}  December 12, 2001\\

\begin{center}\begin{Large}\begin{bf}
No First-Order Phase Transition in the \\ Gross-Neveu Model?\\
\end{bf}\end{Large}\vspace{.75cm}
 \vspace{0.5cm}
Andrej Brzoska\footnote{abrzo@theorie3.physik.uni-erlangen.de} and Michael Thies\footnote{thies@theorie3.physik.uni-erlangen.de}\\
\vskip 0.3cm
{\em Institut f\"ur Theoretische Physik III \\
Universit\"at Erlangen-N\"urnberg \\
Staudtstra{\ss}e 7 \\
D-91058 Erlangen, Germany} \\
\end{center}


%

\vskip 2.0cm

\begin{abstract} \noindent
Within a variational calculation we investigate the role of baryons for the structure of dense matter in the Gross-Neveu model. We construct a trial ground state at finite baryon density which breaks translational invariance. Its scalar potential interpolates between widely spaced kinks and antikinks at low density and the value zero at infinite density. Its energy is lower than the one of the standard Fermi gas at all densities considered. This suggests that the discrete $\gamma_5$ symmetry of the Gross-Neveu model does not get restored in a first order phase transition at finite density, at variance with common wisdom.  
\end{abstract}

\newpage


\section{Introduction}


Quantumchromodynamics (QCD) at finite temperature and chemical potential raises many interesting questions which may be relevant for the physics of the early universe, dense stars or relativistic heavy ion collisions. Since lattice calculations are not yet feasible at $\mu \neq 0$, the QCD phase diagram is still to a large extent speculative, although there is little doubt that the phase structure is quite rich \cite{Rajagopal01}. In view of this situation, it may be of interest to go back to some exactly soluble field theoretic models and try to understand their phase diagram, in particular the structure of dense, baryonic matter (for a recent review, see Ref. \cite{Schoen01}). The simplest such model is perhaps the Gross-Neveu model \cite{Gross74}, $N$ species of massless fermions with a quartic self-interaction, 
\begin{equation}
{\cal L} = \bar{\psi} {\rm i} \gamma^{\mu} \partial_{\mu} \psi + \frac{1}{2}g^2 \left[ (\bar{\psi} \psi)^2 + \lambda (\bar{\psi}
{\rm i} \gamma_5 \psi)^2 \right]\ , \quad \lambda = 0, 1 \ . 
\label{i0}
\end{equation}  
(Flavor indices are suppressed, {\em i.e.}, $\bar{\psi}\psi = \sum_{k=1}^N \bar{\psi}_k \psi_k$ etc.) This model can be solved analytically in the large $N$ limit and shares some properties with QCD, notably asymptotic freedom. It exists in two variants with either discrete ($\lambda=0$) or continuous  ($\lambda=1$) chiral symmetry. For the sake of simplicity, we shall refer to the first model as Gross-Neveu (GN) model, to the second one as 2-dimensional Nambu--Jona-Lasinio (NJL$_2$) model \cite{Nambu60}. Aside from the spectrum which contains massive, unconfined fermions as well as composite mesons and baryons, the phase diagram of these models has also been studied some time ago \cite{Harrington75,Dashen75,Wolff85,Treml89,Barducci95,Sugiyama01}. It is quite non-trivial, including a tricritical point separating lines of first and second order transitions in striking  analogy with recent conjectures about real QCD \cite{Stephanov98} . Here, we shall be mainly interested in the behavior of these models at finite density but zero temperature. There, either the discrete or the continuous chiral symmetry  which are spontaneously broken in the vacuum are believed to get restored in a first order phase transition at the critical density $\rho_{\rm crit}=m/(\pi \sqrt{2})$, with $m$ the dynamical fermion mass in the vacuum. Below this density, matter should consist of droplets of chirally restored phase containing extra fermions surrounded by regions of chirally broken vacuum, much like in the bag model. Above $\rho_{\rm crit}$ the fermions become massless and free. 

This scenario which dates back to the mid 80's was recently challenged in the case of the NJL$_2$ model by one of us \cite{Schoen00}, and we extend our criticism to the GN model in the present work. Our main objection is very simple indeed: The conventional wisdom ignores the fact that these models possess baryons, either massless (NJL$_2$ model) or massive ones (GN model) (see \cite{Schoen01} and references therein). In the low density limit, one would expect that baryonic matter will form widely separated baryons, its energy density satisfying
\begin{equation}
\left. \frac{\partial {\cal E}}{\partial \rho} \right|_{\rho=0} =  M_B \ .
\label{i1}
\end{equation}
This relation is violated by the standard solutions to the GN or NJL$_2$ models. The reason can be traced back to the tacit assumption of unbroken translational invariance inside and outside the droplets. In the case of the NJL$_2$ model, the energetically favored structure of baryonic matter was shown to be a chiral crystal \cite{Schoen00}: Baryons are topological excitations of the pion field, analogous to the Skyrme model in higher dimensions \cite{Skyrme61}, and matter is a kink crystal, the low dimensional analogue of a Skyrme crystal \cite{Klebanov85}. This result can be shown analytically with the help of bosonization techniques; Eq. (\ref{i1}) is then indeed satisfied for the true ground state with $M_B=0$. Strictly speaking, no restoration of chiral symmetry takes place as a function of density, although a rapidly fluctuating condensate may be considered as equivalent to the chirally symmetric phase for most practical purposes. The same argument will apply for the crystalline configuration of the GN model constructed in this work.

The problem in the GN model is more involved, since the bosonization used in Ref. \cite{Schoen00} can only be applied to the light ``pion". Therefore, we cannot present a solution to the finite density GN model (possessing only the massive ``sigma" meson) which would be nearly as complete as for the NJL$_2$ model. We have performed a variational calculation in order to find out whether the standard Fermi gas ground state of baryonic matter in the GN model is unstable against crystallization. 

We finally remind the reader that quite generally, breaking of continuous symmetries and Goldstone bosons are not allowed in 1+1 dimensions \cite{Mermin66,Coleman73}. As is well known, the large $N$ limit suppresses the fluctuations which would otherwise restore the symmetry. This makes it possible to study such effects in 1+1 dimensions as well, provided one works only in leading order in $1/N$, {\em i.e.}, takes the limit $N \to \infty$ at face value. This remark applies both to the continuous chiral symmetry of models like the NJL$_2$ or 't Hooft model \cite{tHooft74} where massless, pion-like excitations appear, and to translational invariance in the GN model, our main theme here.


\section{Construction of a trial wave function}


In the large $N$ limit, the relativistic Hartree-Fock approximation for the ground state becomes exact. It corresponds to a variational calculation where the trial wave functions are all possible Slater determinants; the Hartree-Fock method determines the best fermion single particle orbits. Since we cannot solve this problem fully at this point, we search for the best Slater determinant in a restricted subspace only, generating the single particle orbits from a certain one-parameter family of periodic scalar potentials. These scalar potentials need not be self-consistent since they are only used to restrict the subspace of Slater determinants. By evaluating the expectation value of the exact Hamiltonian $H$ in this trial state and minimizing it with respect to this one parameter, we obtain an upper bound for the ground state energy. If this upper bound is below the energy of the translationally invariant Hartree-Fock solution, we can in this way establish instability of the system with respect to breakdown of translational invariance. In the present section, we outline the construction of our trial wave function. It is motivated primarily by our expectation that matter at low density will tend to form isolated baryons. Another guiding principle is the desire to gain as much analytical insight as possible.

Our starting point is the scalar potential for a single baryon in the Gross-Neveu model. Using units where $m=1$ ($m$ is the physical fermion mass), it is given by \cite{Dashen75a,Pausch91}
\begin{equation}
S(x)=1+y (\tanh \xi_{-} -\tanh \xi_{+} )\ , \quad \xi_{\pm}=yx\pm \frac{1}{2} {\rm artanh}\, y \ . 
\label{gn1}
\end{equation}
The mean-field Dirac Hamiltonian with this potential,
\begin{equation}
h_D = {\gamma_5}\frac{1}{\rm i} \frac{\partial}{\partial x} +\gamma^0 S(x) \ , 
\label{gn1a}
\end{equation}
can be diagonalized analytically. The parameter $y\in [0,1]$ is related to the occupation of the valence level. If $y\ge \sqrt{3}/2$, $S(x)$ crosses zero at the points $ x=\pm R/2 $ with 
\begin{equation}
R=\frac{1}{2y} \ln \left( \frac{2 y^2-1+yw}{2 y^2-1-yw}\right)\ ,
\quad w=\sqrt{4 y^2-3} \ .
\label{gn2}
\end{equation}
Furthermore, we note that as a result of spontaneous breakdown of the discrete $\gamma_5$ symmetry, $S(x)$ and $-S(x)$ are equivalent scalar baryon potentials built on the two degenerate vacua. The corresponding single particle solutions of the Dirac equation can be related by $\psi'(x)={\rm i}\gamma_5 \psi(x)$. This enables us to construct a family of smooth, periodic potentials for which the Dirac equation is again exactly solvable: Choose $S(x)$ from Eq. (\ref{gn1}) in the interval $[-R/2,R/2]$ and extend it to the whole axis via $S(x+R)=-S(x)$. By pasting together pieces of the single baryon potential with alternating signs in this manner, we obtain a continuous and differentiable scalar potential whose period of $2R$ is controlled by the parameter $y$. The shape of the potential changes with the period in a prescribed way, see Fig. 1. In the limit $y\to 1$, the potential corresponds to a succession of equidistant, alternating kinks and anti-kinks at a distance growing logarithmically with $(1-y)$. In the opposite limit $y\to \sqrt{3}/2$, $S$ may be interpreted as arising from strongly overlapping baryons with an almost complete cancellation of their scalar potentials. Only some rapid residual oscillation around zero is left which goes over into the value zero of the chirally restored phase at the endpoint $y=\sqrt{3}/2$. The resulting scalar potential is not only periodic with period $2R$, but also invariant under translations by half a period ($R$) combined with a discrete $\gamma_5$-transformation. In this sense, it may be viewed as discrete version of the ``chiral spiral" of Ref. \cite{Schoen00}. What we have set up here is evidently a kink-antikink crystal, as opposed to the kink crystal characteristic for the ground state of models with continuous chiral symmetry. 

The Dirac equation for a single baryon can be solved in closed form \cite{Dashen75a,Pausch91}. Since we shall be mostly interested in negative energy continuum states in what follows, we recall the well-known results for these particular wavefunctions only. In the interval $[-R/2,R/2]$, one finds (we use a representation where $\gamma^0=-\sigma_1, \gamma^1={\rm i}\sigma_3, \gamma_5=\gamma^0 \gamma^1= -\sigma_2$),
\begin{equation}
\psi_k(x) = \left(  \begin{array}{c} ({\rm i}k-1)({\rm i}k-y \tanh \xi_-) \\
-E(k) ({\rm i}k-y \tanh \xi_+ ) \end{array} \right)  {\rm e}^{{\rm i}kx} \ ,\quad E(k)=\sqrt{k^2+1}\ .
\label{gn3}
\end{equation}
$\psi_k$ and $\psi_{-k}$ are degenerate with eigenvalue $-E(k) $. The general solution of the stationary Dirac equation to $-E(k)$ in this interval is then
\begin{equation}
\psi(x) = A \psi_k(x) + B \psi_{-k}(x) \ .
\label{gn4}
\end{equation} 
Furthermore there exists a discrete valence level at energy $\sqrt{1-y^2}$.

Let us now apply the textbook procedure from solid state physics in order to extend the solution to the whole $x$-axis and determine the spectrum of the Dirac Hamiltonian. In a slight modification of Bloch's theorem, we require the following boundary condition,  
\begin{equation}
\psi(R/2) = {\rm e}^{{\rm i} \delta} {\rm i} \gamma_5 \psi(-R/2) \ .
\label{gn5}
\end{equation}
This guarantees a smooth matching of the wave functions at the borders of different intervals. Inserting the ansatz (\ref{gn4}) and denoting the upper (lower) components of $\psi_k$ by  $\varphi_k$ ($\chi_k$), Eq. (\ref{gn5}) yields the generalized eigenvalue problem
\begin{equation}
M v = {\rm e}^{{\rm i}\delta} N v 
\label{gn6}
\end{equation}
with
\begin{equation}
M=\left(\begin{array}{cc} \varphi_k(R/2) & \varphi_{-k}(R/2)\\ \chi_k(R/2) &  \chi_{-k}(R/2) \end{array} \right) \ ,
\  N=\left(\begin{array}{rr}- \chi_k(-R/2) & -\chi_{-k}(-R/2)\\ \varphi_k(-R/2) &  \varphi_{-k}(-R/2) \end{array} \right) \ ,
\  v=\left(\begin{array}{c} A \\ B \end{array}\right) \ .
\label{gn7}
\end{equation}
Hence the factors  ${\rm e}^{{\rm i}\delta}$ are the eigenvalues of the $2\times 2$-matrix $N^{-1}M$. Evaluation of this matrix using Eq. (\ref{gn3}) shows that it has unit determinant, whereas its trace is given by  
\begin{equation}
{\rm tr} \left(N^{-1}M\right) = 2 X_k\ , 
\label{gn8}
\end{equation}
with
\begin{equation}
X_k=\frac{-E(k)}{k(k^2+y^2)}\left[ (k^2+1-y^2)\sin(kR)+kw\cos(kR)\right] \ .
\label{gn9}
\end{equation}
The characteristic equation then reads
\begin{equation}
( {\rm e}^{{\rm i}\delta})^2 - 2 X_k {\rm e}^{{\rm i}\delta}+1=0 \ .
\label{gn9a}
\end{equation}
For $|X(k)|\leq 1$, the matrix $N^{-1}M$ is unitary, $\delta=\arccos X_k$ is real and the corresponding level lies inside an allowed band. For $|X(k)|>1$, $\delta$ is imaginary and we set $\delta={\rm i}\zeta$. The eigenvalues of $N^{-1}M$ are now ${\rm e}^{\pm \zeta}$ with wave functions blowing up in one direction --- the corresponding state is forbidden and lies inside a band gap. The diagonalization of Eq. (\ref{gn6}) also determines the ratio $A/B$ of the coefficients introduced in Eq. (\ref{gn4}); one finds
\begin{equation}
\frac{A}{B}=\frac{1+{\rm i}k}{1-y^2}\left(Y_k \pm \sqrt{Y_k^2-(1-y^2)^2/E(k)^2}\right) \ ,
\label{gn10}
\end{equation}
with
\begin{equation}
Y_k=kw \sin(kR) - (k^2+1-y^2)\cos (kR) \ .
\label{gn11}
\end{equation}

The two solutions correspond to the two degenerate states with energy $-E(k)$. This procedure yields both the spectrum of the Dirac Hamiltonian and the single particle wave functions up to the normalization for which we shall use a box normalization later on (cf. Sect. 3). Here, let us briefly illustrate the resulting band structure of the potential used to generate our trial wave function.  In Fig. 2, we show an example of the single particle spectrum, comparing it to the one corresponding to the interacting vacuum (or, equivalently, free massive fermions with $m=1$). It exhibits the expected band structure. Since all eigenvalues still occur in pairs $\pm E(k)$, the Dirac sea inherits this band structure, a novel feature as compared to a non-relativistic solid. Fig. 3 displays the lowest few bands as function of the parameter $y$. The left-hand part of this figure (corresponding to small lattice spacing $2R$) can be understood perturbatively, since the scalar potential is weak (cf. Fig. 1). Using degenerate perturbation theory, one easily finds the following position and width of the gaps, expressed in terms of  $\eta=y-\sqrt{3}/2$, 
\begin{equation}
E_{\rm gap} \approx \frac{(2n+1)\pi}{3^{(1/4)} 8 \sqrt{\eta}}\left(1-\frac{\eta}{2\sqrt{3}}\right) \ , \qquad 
\Gamma_{\rm gap} \approx \frac{64 \sqrt{3} \eta}{\pi^3 (2n+1)^3} \ , \qquad n=0,1,2...
\label{gn13}
\end{equation}
On the right-hand side of Fig. 3 (large lattice spacing), all gaps move into the mass gap and we recover the spectrum of the Callan-Coleman-Gross-Zee kink \cite{Dashen75a}, including the valence state at zero energy. The lowest allowed band obviously develops out of the valence level of the single baryon (which moves to zero energy as $y\to 1$), see the dashed curve in Fig. 3. Finally, in Fig. 4, we illustrate the band structure together with the corresponding scalar potential, for three values of $y$. Case a) features well separated baryons, weak tunneling and hence a  narrow valence band. In case b), the band width is comparable to the depth of scalar potential. In case c), the valence band has become much wider than the amplitude of the scalar potential. Fermions can move freely through the whole crystal and baryons have lost their identity. By varying the parameter $y$, we let the system decide which option it prefers at a given density.  


\section{Computing the regularized ground state energy}


In our variational calculation of the ground state, we proceed as follows: We take the periodic scalar potential from the preceding section and fill the single particle states until the desired average baryon density is reached. Next, we evaluate the expectation value of the exact Hamiltonian in this trial state and minimize with respect to  $y$, our only variational parameter. At the minimum, there will be a unique relationship between lattice spacing $2R$ (or, equivalently, $y$) and baryon density. In the case of the single baryon, $y$ was linked to the occupation fraction $\nu = n/N$ of the valence level via $y=\sin (\pi \nu/2)$.

Evaluation of the ground state energy cannot be done naively for several reasons. First, there is a trivial quadratic divergence in the kinetic energy due to the Dirac sea. This will be removed by subtracting the ground state energy of a free Fermi gas of corresponding density. Then, there are logarithmic divergences both in the kinetic and in the potential energy which require the usual renormalization of the theory at zero density. Moreover, due to the gap structure of the Dirac sea, a direct calculation in the continuum would be technically difficult. We therefore enclose the system in a box of size $L$ (an integer multiple of the lattice spacing,  $L=2RK$) and require antiperiodic boundary conditions, thereby discretizing the spectrum of $h_D$. The size of the box will be made large enough so that our results do not depend on the IR-regulator $L$. The phase factor ${\rm e}^{{\rm i}\delta}$ in Eq. (\ref{gn5}) then has to satisfy
\begin{equation}
(-{\rm e}^{2{\rm i}\delta})^K=-1 \ , \quad \delta=p_n R+\frac{\pi}{2} \ , \quad p_n=\frac{\pi}{L}(2n+1)
\label{gn13a}
\end{equation}
with $n$ an integer. Solving the characteristic equation (\ref{gn9a}) yields
\begin{equation}
{\rm e}^{{\rm i}\delta} = X_k \pm {\rm i} \sqrt{1-X_k^2} 
\label{gn13b}
\end{equation}
or, equivalently,
\begin{equation}
- \sin(p_nR)=X_k \ .
\label{gn14}
\end{equation}
A solution of Eq. (\ref{gn14}) for $k$ then yields the single particle energies $\pm E(k)$ corresponding to the discrete $p_n$. 

To reach a certain baryon density, one has to fill all negative energy states plus a number of positive energy states. ``Antimatter" on the other hand is obtained by leaving a number of negative energy levels unoccupied. Since the energy density is identical in both cases, we have found it more convenient to lower the level of the Dirac sea rather than raise it. This is the reason why \  --- as mentioned above ---\ we need to consider only negative energy single particle states below the ``Fermi surface". We denote the single particle wave functions by  $\psi_{n,i}$, where $n$ refers to the energy  and $i=1,2$ labels the two degenerate solutions.

Taking the expectation value of $H$ in our trial state and invoking the large $N$ limit yields the following expression for the ground state energy per flavor,  
\begin{equation}
E=T+V
\label{b1}
\end{equation}
with
\begin{equation}
T= \sum_n^{occ} \sum_{i=1}^2 2K \int_{-R/2}^{R/2} {\rm d}x\, 
 \frac{1}{{\cal N}_{n,i}} \psi_{n,i}^{\dagger} \gamma_5 \frac{1}{\rm i}
\frac{\partial}{\partial x} \psi_{n,i} \ ,
\label{b2}
\end{equation}
\begin{equation}
V=-Ng^2 K \int_{-R/2}^{R/2} {\rm d}x\, \langle \bar{\psi}\psi \rangle^2\ , \qquad
\langle \bar{\psi}\psi \rangle = \sum_n^{occ} \sum_{i=1}^2 \frac{1}{{\cal N}_{n,i}} \bar{\psi}_{n,i}\psi_{n,i}\ .
\label{b5}
\end{equation}
The normalization factors are
\begin{equation}
{\cal N}_{n,i} = 2 K \int_{-R/2}^{R/2} {\rm d}x\, \psi_{n,i}^{\dagger} \psi_{n,i}\ .
\label{b3}
\end{equation}

The kinetic energy has both quadratic and  logarithmic divergences. Let us isolate the divergences by adding and subtracting the asymptotic form ($n\to \infty$) of the terms in Eq. (\ref{b2}),
\begin{equation}
T=T_{\rm conv}+T_{\rm div} \ ,
\label{b6}
\end{equation}  
with
\begin{eqnarray}
T_{\rm conv} &=& \sum_n^{occ} \left\{ \sum_{i=1}^2  2K \int_{-R/2}^{R/2} {\rm d}x\, 
\frac{1}{{\cal N}_{n,i}} \psi_{n,i}^{\dagger} \gamma_5 \frac{1}{\rm i}
\frac{\partial}{\partial x} \psi_{n,i} +2 p_n-\frac{1}{p_n}\left(1-\frac{2w}{R}\right)\right\}\ , 
\nonumber \\
T_{\rm div} & =& \sum_n^{occ}\left\{- 2 p_n+\frac{1}{p_n}\left(1-\frac{2w}{R}\right)\right\}\ .
\label{b8}
\end{eqnarray}

Here, $p_n$ are the momenta of the free fermions introduced in Eq. (\ref{gn13a}). One can derive this asymptotic form using the known single particle wave functions of Sect. 2. The potential energy has a logarithmic divergence which can be exposed as follows. First, split the condensate into regular and singular pieces,
\begin{equation}
\langle \bar{\psi} \psi \rangle = \langle \bar{\psi} \psi \rangle_{\rm conv}+\langle \bar{\psi}\psi \rangle_{\rm div} \ ,
\label{b10}
\end{equation}
with
\begin{eqnarray}
\langle \bar{\psi} \psi \rangle_{\rm conv} &=& \sum_n^{occ} \sum_{i=1}^2 \frac{1}{{\cal N}_{n,i}} \bar{\psi}_{n,i}\psi_{n,i}
+ \frac{1}{L} \sum_n^{occ} \left( \frac{2S}{p_n} \right) \ , \nonumber \\
\langle \bar{\psi} \psi \rangle_{\rm div} &=& - \frac{1}{L} \sum_n^{occ} \left( \frac{2S}{p_n} \right) \ .
\label{b12}
\end{eqnarray}

Once again, the explicit form of the single particle wave functions has been used to derive Eq. (\ref{b12}), where $S$ is the scalar potential described above. As usual, the gap equation at zero density is used to renormalize the coupling constant,
\begin{equation}
\frac{1}{Ng^2} = \frac{2}{L} \sum_n^{vac} \frac{1}{E(p_n)}\ .
\label{b13}
\end{equation}
Here, the summation over $n$ is according to the vacuum occupation, as opposed to the occupation for baryonic matter which enters the sums in Eqs. (\ref{b2},\ref{b5}). The UV-cutoff $\Lambda$ has to be the same in both sums. We then insert the decomposition (\ref{b10}) into Eq. (\ref{b5}). Since $Ng^2$ vanishes like $(\ln \Lambda)^{-1}$, we may drop the $\langle \bar{\psi}\psi \rangle_{\rm conv}^2$ term and have to keep only the two remaining terms,
\begin{equation}
V=V_1 + V_2
\label{b15}
\end{equation}
with
\begin{eqnarray}
V_1 &=& - Ng^2 K\int_{-R/2}^{R/2} {\rm d}x \, 2 \langle \bar{\psi} \psi \rangle_{\rm conv} \langle \bar{\psi} \psi
\rangle_{\rm div} \ , \nonumber \\
V_2 & = &  - Ng^2 K\int_{-R/2}^{R/2} {\rm d}x \,  \langle \bar{\psi} \psi
\rangle_{\rm div}^2  \ .
\label{b16}
\end{eqnarray}

Using Eqs. (\ref{b12}) and (\ref{b13}) and performing the limit $\Lambda \to \infty$, $V_1$ reduces to
\begin{equation}
V_1= 2 K \int_{-R/2}^{R/2} {\rm d}x\, \langle \bar{\psi}\psi \rangle_{\rm conv} S \ .
\label{b17}
\end{equation}

$V_2$ is given by
\begin{equation}
V_2= - \frac{ \left(\sum_n^{occ}\frac{1}{p_n} \right)^2}{\left(\sum_n^{vac}\frac{1}{E(p_n)}\right)} \frac{1}{R} \int_{-R/2}^{R/2} {\rm d}x \, S^2\ .
\label{b18}
\end{equation}

The integral can be performed in closed form with the compact result
\begin{equation}
\int_{-R/2}^{R/2} {\rm d}x \, S^2\ =(R-2w) \ .
\label{b18a}
\end{equation}
Assuming  the cutoff  $\Lambda$ to be large then yields
\begin{equation}
V_2 = - \left\{\sum_n^{occ} \frac{1}{p_n} -\left( \sum_n^{vac} \frac{1}{E(p_n)} - \sum_n^{occ} \frac{1}{p_n} \right) \right\} \left( 1-\frac{2w}{R}
\right) \ .
\label{b19}
\end{equation}

If we add up $T_{\rm div}$, Eq. (\ref{b8}) and $V_2$, Eq. (\ref{b19}), and eliminate the trivial quadratic divergence by subtracting the free Fermi gas kinetic energy at the same density, we obtain the UV finite result
\begin{equation}
T_{\rm div}+V_2 + \sum_n^{occ} 2 p_n = \left( 1-\frac{2w}{R} \right) \left(
 \sum_n^{vac} \frac{1}{E(p_n)} - \sum_n^{occ} \frac{1}{p_n} \right) \ .
\label{b20}
\end{equation}

After some rearrangement, the total regularized energy can be cast into the form
\begin{eqnarray}
E_{\rm reg} &=& 
\sum_n^{occ} \left\{ \sum_{i=1}^2  2 K \int_{-R/2}^{R/2} {\rm d}x  
\frac{1}{{\cal N}_{n,i}} \psi_{n,i}^{\dagger} \gamma_5 \frac{1}{\rm i}
\frac{\partial}{\partial x} \psi_{n,i} +2 p_n-\frac{1}{p_n}\left(1-\frac{2w}{R}\right)\right\} \nonumber \\
& & + 2 K \int_{-R/2}^{R/2} {\rm d}x \, S \left\{\sum_n^{occ} \sum_{i=1}^2 \frac{1}{{\cal N}_{n,i}} \bar{\psi}_{n,i}\psi_{n,i}
+ \frac{S}{L}\left( \sum_n^{vac} \frac{1}{E(p_n)} + \sum_n^{occ} \frac{1}{p_n} \right) \right\} \ . 
\label{b21}
\end{eqnarray}

Invoking the fact that the $\psi_{n,i}$ are eigenfunctions of the Dirac Hamiltonian (\ref{gn1a}) with eigenvalues $-E(k_n)$ as well as the result (\ref{b18a}), this expression can further be simplified to
\begin{equation}
E_{\rm reg}=-2 \sum_n^{occ} \left( E(k_n)-p_n\right)
+    \left(1-\frac{2w}{R} \right) \sum_n^{vac} \frac{1}{E(p_n)}\ .
\label{b22}
\end{equation}

The $k_n$ are the single particle momenta to the corresponding discrete $p_n$, obtained via the transcendental equation (\ref{gn14}).

Remarkably, the total energy has been expressed entirely in terms of the single particle energies and known functions of $y$. That the sum in $E_{\rm reg}$ is UV finite can now be confirmed by means of the asymptotic behavior
\begin{equation}
k_n   \approx   p_n - \frac{w}{R p_n} \qquad (n\to \infty) \ ,
\label{b23}
\end{equation}
which in turn follows from the transcendental equation (\ref{gn14}). Summarizing, the calculation of the variational energy has been reduced to solving a transcendental equation for $k_n$, Eq. (\ref{gn14}),  and performing a convergent sum, Eq. (\ref{b22}). The baryon density is adjusted via the occupation of the single particle states. There is only one variational parameter $y$ --- it controls the spatial period of the scalar potential underlying our trial wave function.


\section{Results of the variational calculation}


We have performed calculations along the lines outlined above and evaluated the ground state energy of baryonic matter with our variational ansatz, the kink-antikink crystal. The size of the box $L$ was increased until the results were stable which required values of $L\approx 300-1600$ (in units where $m=1$) increasing with density. The sum to be performed is UV finite so that no cutoff is required. To get precise results, we had to take into account single particle orbits with energies up to $|E| \approx 50-100$ typically.

We first summarize our findings concerning the variational parameter $y$. The result is extremely simple and can be stated as follows: At all densities considered, the best choice of $y$ was the one which yielded a fully occupied valence band in baryonic matter (or empty in ``antimatter"), cf. the shading in Fig. 4. With hindsight, this is rather plausible: Around each gap, the levels are pushed apart symmetrically due to the interaction (in first order perturbation theory). When summing over occupied states the main effect cancels out. Only at the Fermi surface the filling is asymmetric and the system can take advantage of the level shifts to lower its energy. This picture was confirmed by studying in detail where the bulk of the energy difference came from. It was indeed the region in the vicinity of the Fermi surface. The optimal value of $y$ translates into the following relation between baryon density and baryon radius (or lattice spacing),
\begin{equation}
\rho_B=\frac{1}{2R} \ .
\label{b22a}
\end{equation} 
By this relation, the different values for $y$ in Fig. 4 correspond to baryon densities a) $\rho=0.12\,$, b) $\rho=0.3\,$, and c) $\rho=1.0\,$. Each potential well may be associated with one baryon, although for high densities fermions are not confined to one well, cf. Fig. 4.

With this optimal choice of $y$ we get a unique answer for the best ground state energy as a function of density. It is shown in Fig. 5, together with the energies of the free Fermi gas and the energy for the standard ``mixed phase" solution. We recall that in the latter phase, the extra fermions are put into droplets of chirally restored phase, like in the bag model, whereas the vacuum in between the baryons remains in the broken phase. The standard first order phase transition occurs at $p_f=1/\sqrt{2}$, or $\rho\approx 0.225$, as indicated by the dot in Fig. 5. Our result is always below these other two curves. At small densities, it perfectly matches the expected slope, Eq. (\ref{i1}), with $M_B=2/\pi$, the mass of a single baryon with fully occupied valence level. This is no surprise since our variational ansatz was designed to have this limit correctly built in, but it is a good test of the whole calculation, including the regularization and renormalization procedure. In our ansatz, we had to assume that the distance between two baryons equals the baryon diameter in order to generate a smooth periodic potential by gluing together pieces of the single baryon potentials. This is certainly the right thing to do at very low densities. There, the extra fermions are concentrated in the kinks and anti-kinks. The interaction between kinks and anti-kinks is repulsive, therefore they will tend to spread equally; there is no distinction between left and right neighbours. It is more surprising that such an ansatz is also capable of lowering the energy at higher densities, even at densities where the standard picture predicts a transition to the chirally restored phase. Thereby, this standard first-order phase transition in the GN model can be ruled out, because our variational upper bound for the ground state energy excludes the chirally symmetric solution. In a sense, the mixed phase picture may be regarded as a crude attempt to describe baryons. It misses some important binding effects which survive the large $N$ limit \cite{Witten79}, as manifested through the existence of single baryons in the spectrum of the GN model.

We have pushed the calculation up to higher densities without any indication that the curves shown in Fig. 5 will cross each other, see Fig. 6. The energy difference becomes tiny and the volume one needs to get the same accuracy becomes huge, so that it gets increasingly difficult to get precise numerical results. Our trial state definitely becomes worse with increasing density, as can be judged by comparing the chiral condensate $\langle \bar{\psi}\psi\rangle$ computed from our wave functions with the input scalar potential. Hence the true ground state must be lower in energy at high densities than suggested by our variational calculation.

Finally, in Fig. 7, we illustrate the spatial dependence of the baryon density for the three values of $y$ already used in Fig. 4. Unlike in the NJL$_2$ model where the baryon density was constant, here it also reflects the periodicity of the scalar potential.

Needless to say, it would be of quite some theoretical interest to find the truly self-consistent periodic Hartree-Fock potential for the GN model at finite density. This may be the prerequisite for revising the phase diagram of the GN model in the whole $(T,\mu)$ plane.

\bibliographystyle{unsrt}

\newpage

\begin{figure}[t]
 \begin{center}
  \epsfig{file=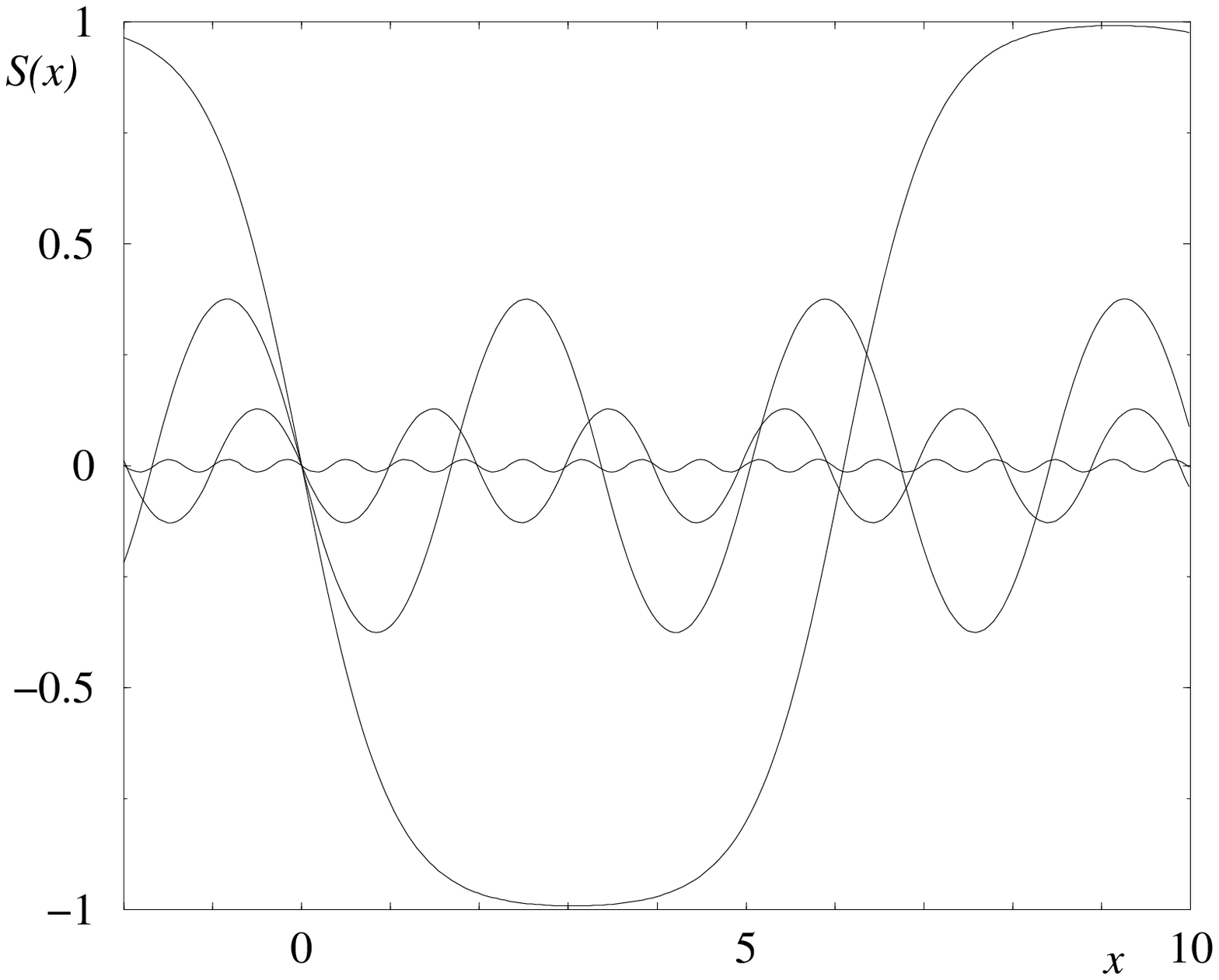, width=8cm, height=7cm}
   \caption{Periodic scalar potentials with half period $R=0.33\,$, $0.99\,$, $1.68\,$, and $6.10\,$.}
  \end{center}
\end{figure}

\begin{figure}[t]
 \begin{center}
  \epsfig{file=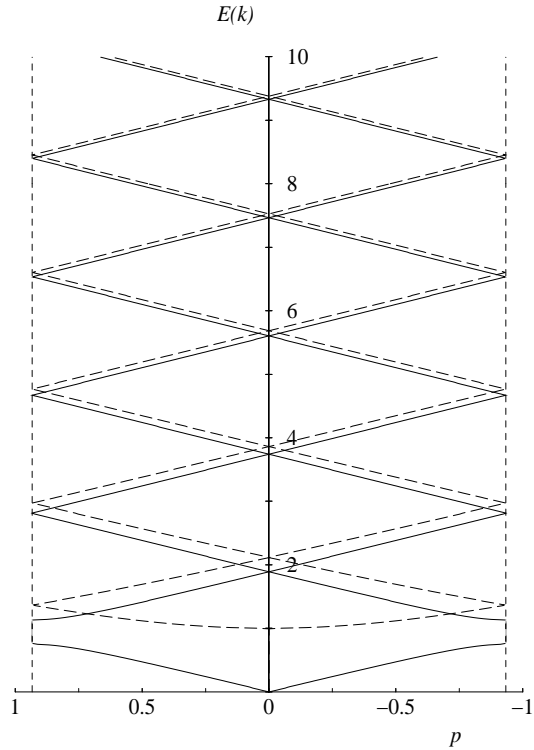, width=7cm, height=10cm}
   \caption{Single particle spectrum in a reduced zone scheme as function of the Bloch momentum $p=\frac{1}{R}\left(\delta-\frac{\pi}{2}\right)$, restricted to the first Brillouin zone $\left[-\frac{\pi}{2R},\frac{\pi}{2R}\right]$. Solid line: single particle energies obtained from the transcendental equation $\delta(k)=\arccos X_k$ for $y=0.95\,$; dashed line: single particle energies in the interacting vacuum.}
  \end{center}
\end{figure}

\begin{figure}[t]
 \begin{center}
  \epsfig{file=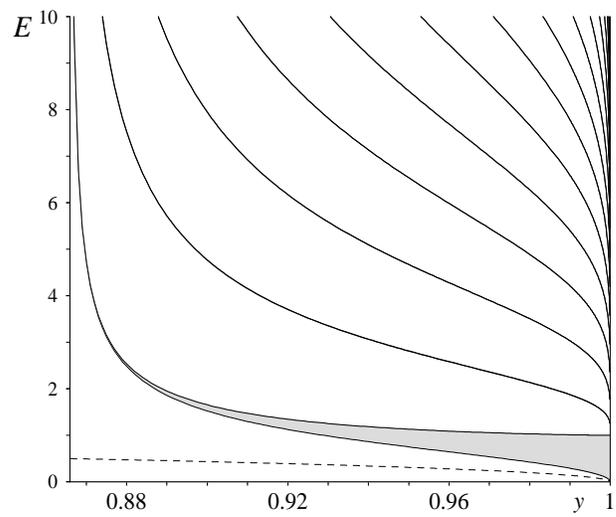, width=8cm, height=7cm}
   \caption{The lowest band gaps (solid lines and shaded area) as function of $y$; the white regions are the allowed bands. The dashed line is the position of the discrete valence level of the single baryon.}
  \end{center}
\end{figure}

\begin{figure}[t]
  \begin{center}
    a)\quad\raisebox{-4.75cm}{\epsfig{file=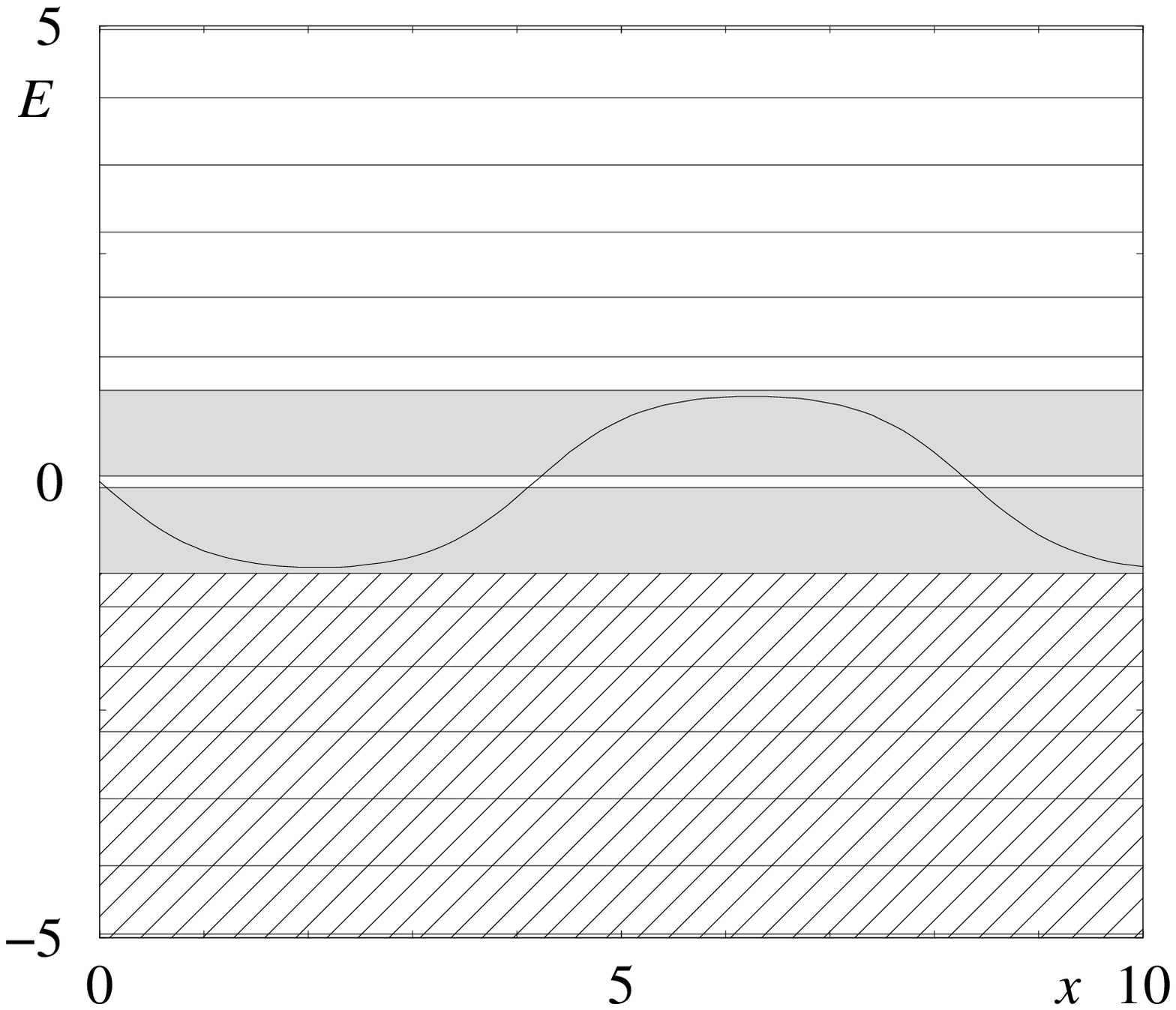, width=6cm, height=5.25cm}}\\[0.3cm]
    b)\quad\raisebox{-4.75cm}{\epsfig{file=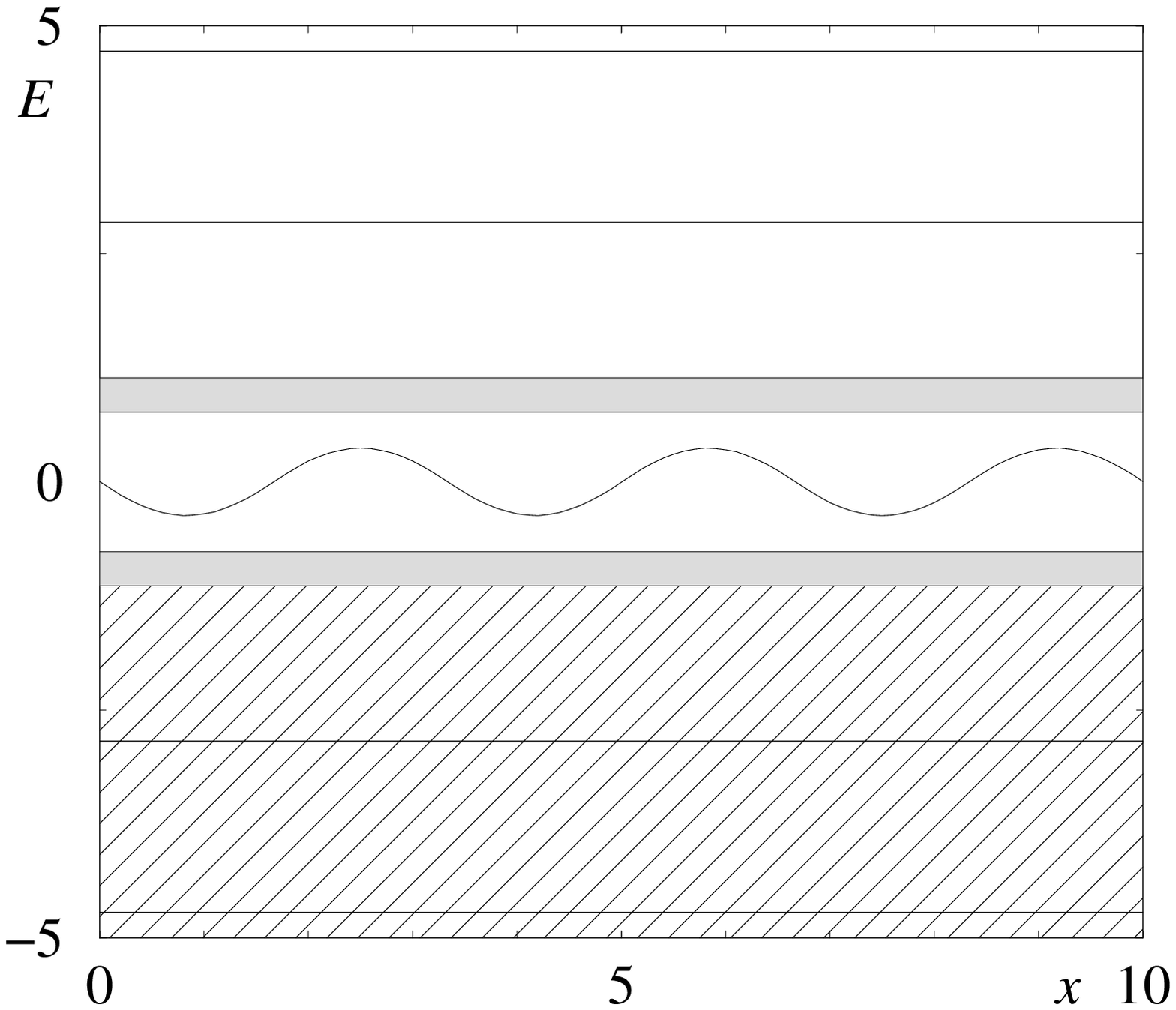, width=6cm, height=5.25cm}}\\[0.3cm]
    c)\quad\raisebox{-4.75cm}{\epsfig{file=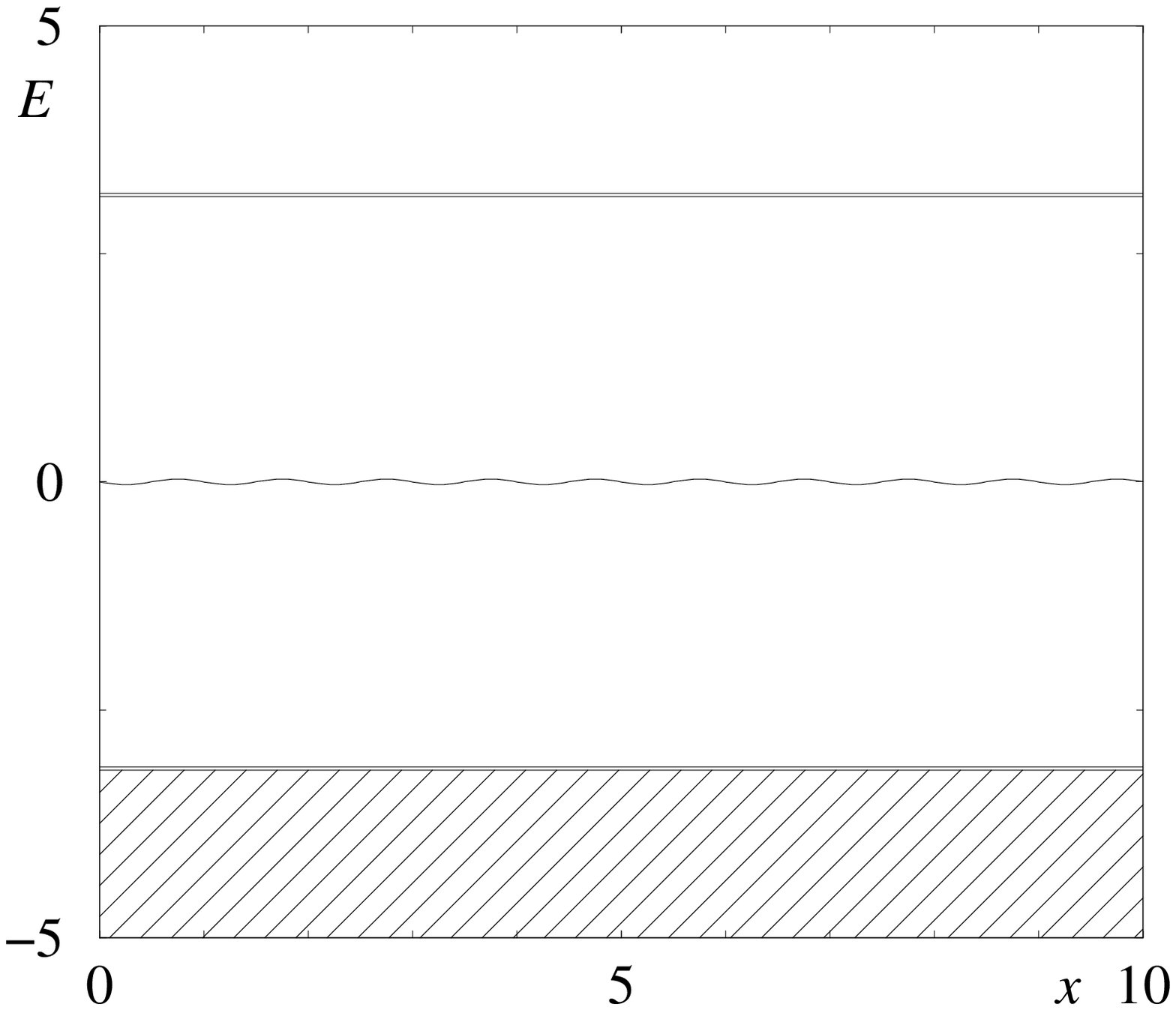, width=6cm, height=5.25cm}}
    \caption{The lowest band gaps (straight, solid lines and shaded areas) in comparison to the corresponding scalar potential (wiggly line), for a) $R=4.16\,$, b) $R=1.66\,$, and c) $R=0.5\,$. The hatched regions are the occupied levels in the antimatter crystal.}
\end{center}
\end{figure}

\begin{figure}[t]
 \begin{center}
  \epsfig{file=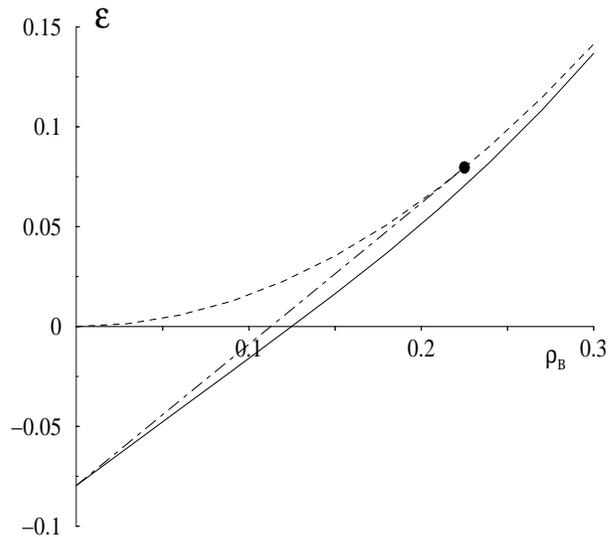, width=8cm, height=7cm}
   \caption{Energy density per flavor as function of the baryon density $\rho_B$. Dashed line: non-interacting Fermi gas; dash-dotted line: standard mixed phase configuration, with the first order phase transition at the dot; solid line: variational upper bound for the kink-antikink crystal.}
  \end{center}
\end{figure}

\begin{figure}[t]
 \begin{center}
  \epsfig{file=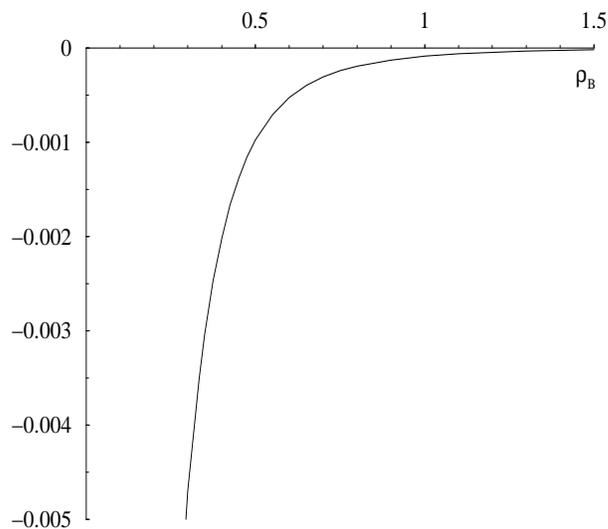, width=8cm, height=7cm}
   \caption{Difference of the energy densities per flavor between the crystal and the trivial Fermi gas solutions. The negative sign shows that the crystal is energetically favored at all densities considered.}
  \end{center}
\end{figure}

\begin{figure}[t]
 \begin{center}
   a)\quad\raisebox{-4.75cm}{\epsfig{file=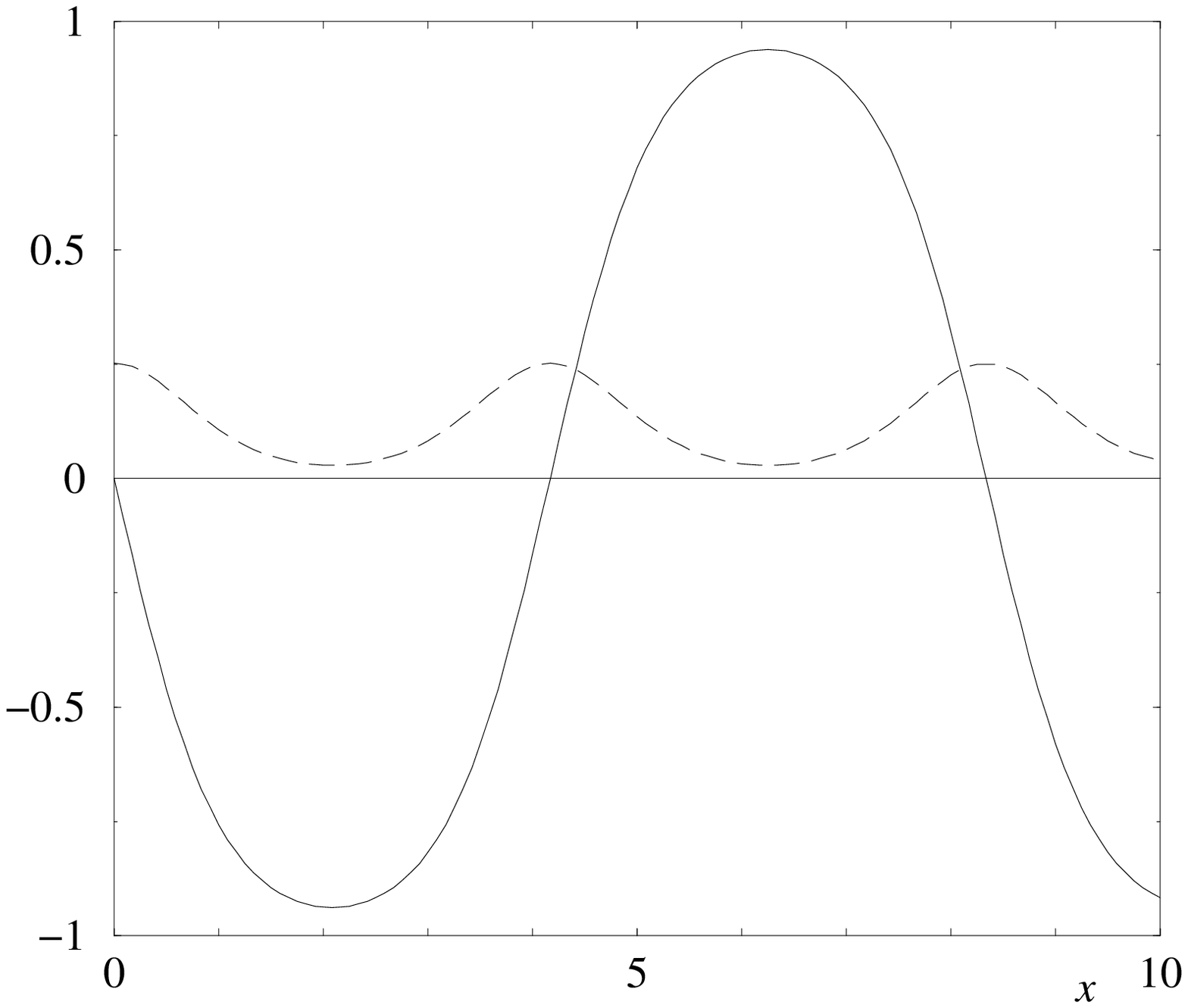, width=6cm, height=5.25cm}}\\[0.3cm]
  b)\quad\raisebox{-4.75cm}{\epsfig{file=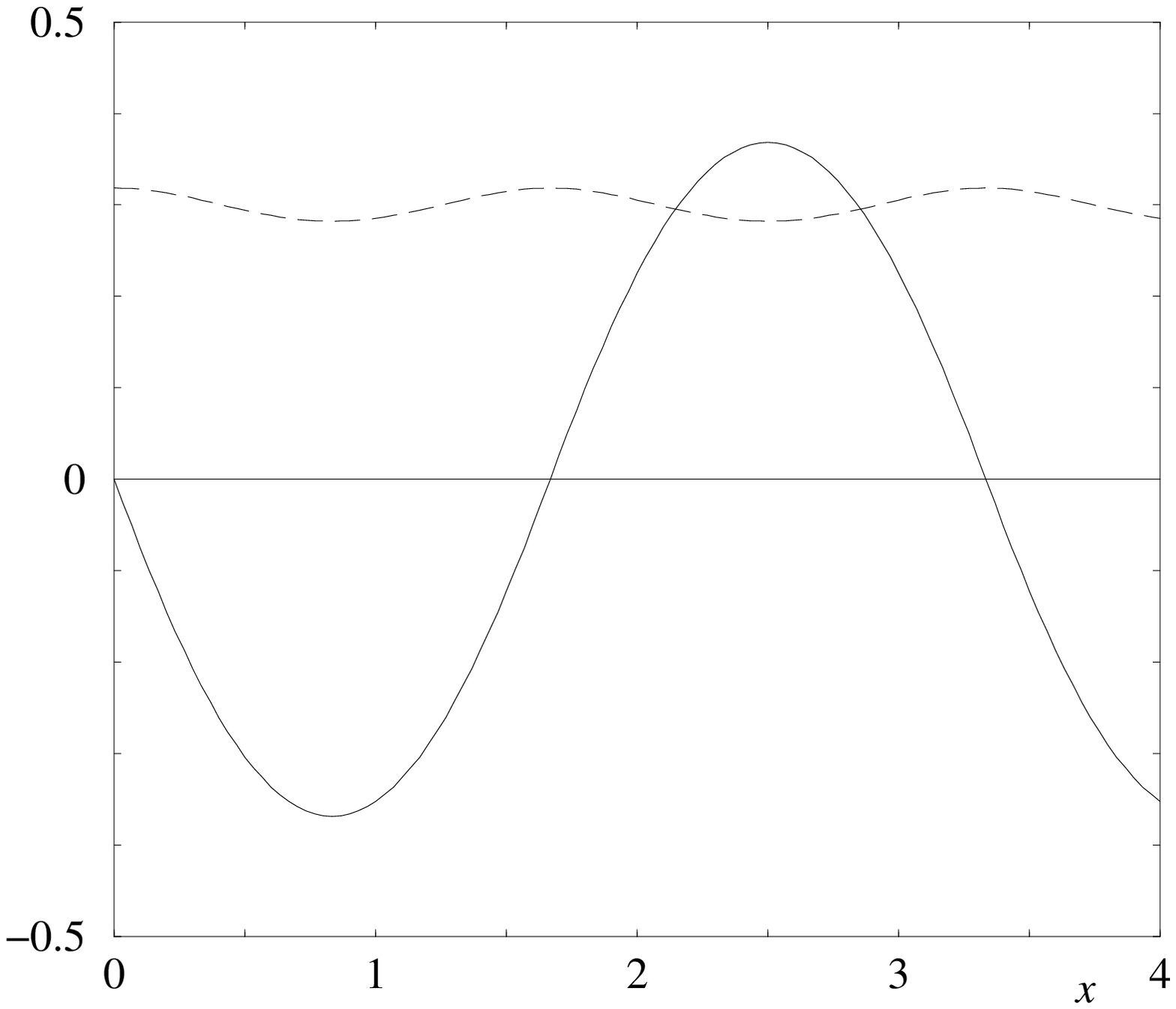, width=6cm, height=5.25cm}}\\[0.3cm]
  c)\quad\raisebox{-4.75cm}{\epsfig{file=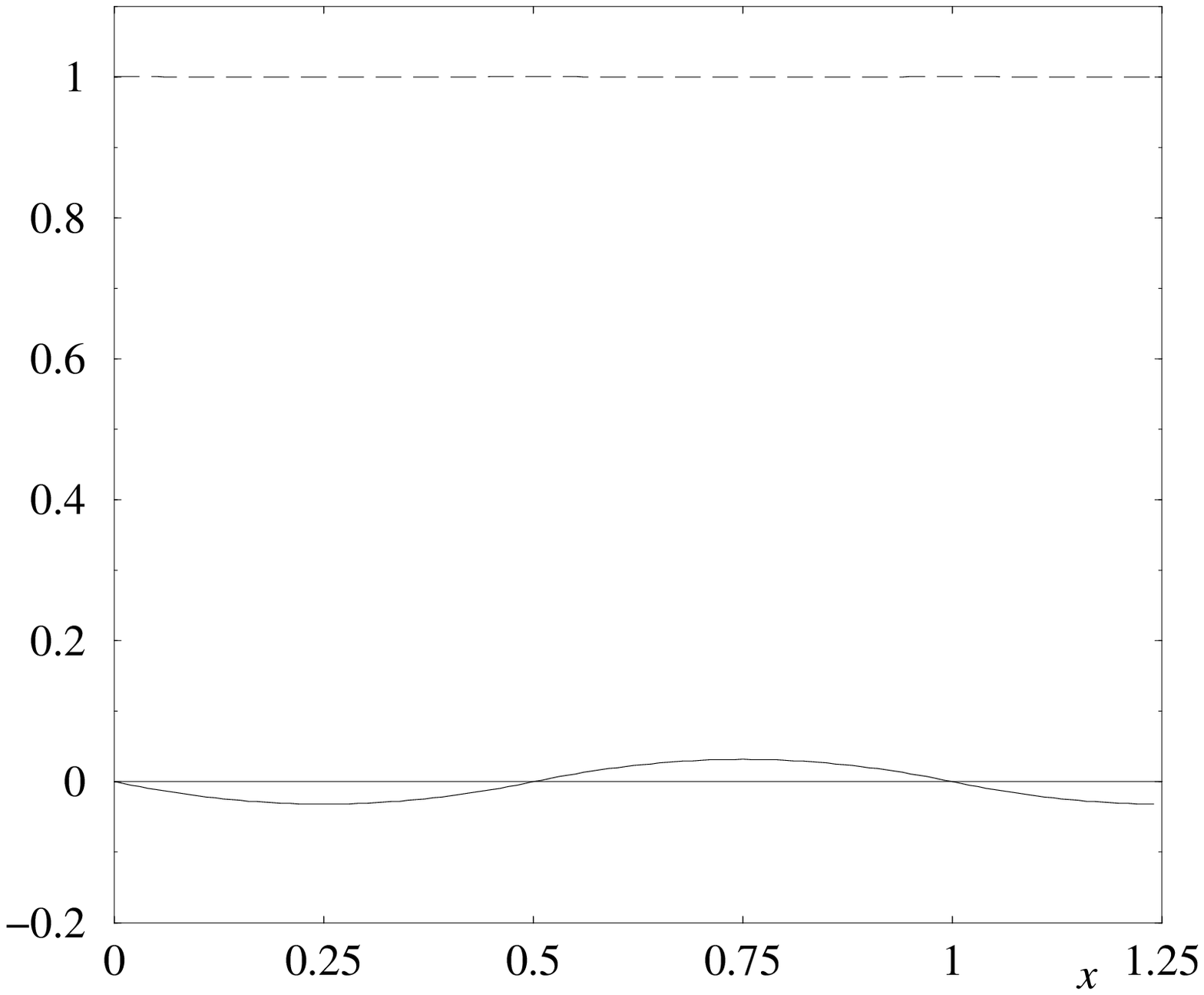, width=6cm, height=5.25cm}}
   \caption{Spatial dependence of the baryon density (dashed), compared to the scalar potential (solid), for mean baryon densities a) $\rho=0.12\,$, b) $\rho=0.3\,$, and c) $\rho=1.0\,$. For $\rho=1$ the oscillation of the baryon density is too weak to be resolved.}
  \end{center}
\end{figure}

\end{document}